\documentclass[11pt,a4,epsf,fleqn]{article}
\usepackage[utf8]{inputenc}
\usepackage{newunicodechar}
\newunicodechar{−}{\textminus}
\usepackage{verbatim}
\usepackage{eurosym,rotating,float}
\usepackage{wrapfig, framed, caption}
\usepackage{caption}
\usepackage{graphicx}
\usepackage{multirow}
\usepackage{booktabs}
\usepackage{soul}
\usepackage{todonotes}
\usepackage{mathcomp}
\usepackage{textcomp}
\usepackage[para,online,flushleft]{threeparttable}
\usepackage{longtable}
\usepackage{bm}
\usepackage[sumlimits,fleqn]{amsmath}
\usepackage{array}
\newcolumntype{K}[1]{>{\centering\arraybackslash}m{#1}}
\usepackage[sort&compress]{natbib}
\usepackage[final]{pdfpages}
\usepackage{arydshln}
\usepackage[bookmarksopen,bookmarksnumbered,colorlinks,linkcolor={blue},citecolor={blue}]{hyperref}
\usepackage{cleveref}[2012/02/15]
\crefformat{footnote}{#2\footnotemark[#1]#3}
\usepackage{color}
\usepackage[displaymath,mathlines,pagewise]{lineno}
\newcolumntype{P}[1]{>{\centering\arraybackslash}p{#1}}
\newcolumntype{L}{>{\centering\arraybackslash}m{3cm}}
\newcolumntype{M}[1]{>{\centering\arraybackslash}m{#1}}
\usepackage{lscape}
\usepackage{pdflscape}
\usepackage[T1]{fontenc}
\usepackage{adjustbox}
\usepackage{mathtools}
\usepackage{enumitem}
\usepackage{soul}
\soulregister\ref7

\soulregister\pageref7
\soulregister\citep7
\soulregister\citet7
\soulregister\citealp7
\soulregister\citealt7
\usepackage[most]{tcolorbox}

\usepackage{xcolor}
\setlist[description]{leftmargin=\parindent,labelindent=\parindent-0.2cm}
\setlist{noitemsep}
\DeclareMathSymbol{\mathdblquotechar}{\mathalpha}{letters}{`"}

\newcommand{\mathdblquote}{\mathtt{\mathdblquotechar}}
\begingroup\lccode`~=`"\lowercase{\endgroup
  \let~\mathdblquote
}
\AtBeginDocument{\mathcode`"="8000 }

 \usepackage{listings}
\begin{document}
\title{Combine and conquer: model averaging for out-of-distribution forecasting}
\author{Stephane Hess\thanks{TU Delft \& University of Leeds, s.hess@leeds.ac.uk} \and Sander van Cranenburgh\thanks{TU Delft, S.vanCranenburgh@tudelft.nl}}

\maketitle

\begin{abstract}
Travel behaviour modellers have an increasingly diverse set of models at their disposal, ranging from traditional econometric structures to models from mathematical psychology and data-driven approaches from machine learning. A key question arises as to how well these different models perform in forecasting, especially when considering trips of different characteristics from those used in estimation, i.e. out-of-distribution prediction, and whether better predictions can be obtained by combining insights from the different models. We focus on trip distance as a key example of a variable where the application context might go beyond the estimation data. Across two case studies, we show that while data-driven approaches excel in predicting mode choice for trips within the distance bands used in estimation, beyond that range, the picture is fuzzy. To leverage the relative advantages of the different model families and capitalise on the notion that multiple `weak' models can result in more robust models, we put forward the use of a model averaging approach that allocates weights to different model families as a function of the \emph{distance} between the characteristics of the trip for which predictions are made, and those used in model estimation. Overall, we see that the model averaging approach gives larger weight to models with stronger behavioural or econometric underpinnings the more we move outside the interval of trip distances covered in estimation. Across both case studies, we show that our model averaging approach obtains improved performance both on the estimation and test data, and crucially also when predicting mode choices for trips of distances outside the range used in estimation. While our initial proof of concept focuses on trip distance as a single trip characteristic to quantify the degree of an observation being out-of-distribution, we also provide initial insights into extending this to a multivariate context, using a Gower distance metric.\\ 

\end{abstract}

\section{Introduction}
The travel behaviour modelling literature has focused extensively on two sorts of models, namely models for inference (henceforth inference models) and models for forecasting (henceforth forecasting models). Inference models aim to understand current travel behaviour (e.g. to recover the value of travel time), while forecasting models aim to predict future travel behaviour in new settings (e.g. due to changes in transport policies, such as toll roads and fuel levies). 

Inference models and forecasting models are evaluated differently by analysts. When building an inference model, the focus is on ensuring that the model generates behaviourally plausible insights into explanatory factors and their relative impacts. In contrast, when building a forecasting model, the emphasis is on the model's ability to generate accurate predictions of behaviour. Importantly, forecasting involves predicting behaviour in new settings, meaning that the models have to generalise out-of-sample. This may relate, for example, to specific combinations of time and cost that were not observed in the estimation data. In addition, some prediction settings may lie outside the range of attribute values on which the models were estimated. This could involve higher costs, longer travel times, or, more intuitively, trips over longer distances\footnote{We do not consider issues of spatial or temporal transferability \citep[see e.g.][]{Fox_Daly_Hess_Miller_2014}, nor the introduction of new modes or attributes. Instead, we focus on situations where attribute levels, or combinations thereof, fall outside the distribution observed in the estimation data. Our analysis is further restricted to continuous attributes, or categorical attributes with pre-determined levels, for which the impact of new attribute levels can be evaluated based on the estimated model parameters.}. In such cases, it is important for forecasting models to also generalise out-of-distribution.

Travel behaviour analysts are increasingly interested in a more diverse set of models, covering not just econometric structures, but also models from mathematical psychology \citep[e.g.][]{hancock2018accumulation,HANCOCK2020165}, and increasingly machine learning techniques \citep[e.g.][]{alwosheel2021did, wong2021reslogit,sfeir2022gaussian,yan2025enhancing,WANG2020234}. This is especially true given the growing reliance on highly granular data on travel behaviour, for example that collected using GPS tracking \citep[e.g.][]{liu2024enhancing,TSOLERIDIS2025105289}. With the typical model evaluation approach focusing on the ability of models to replicate the behaviour in the empirical data as well as possible, it should then come as no surprise that researchers in the travel behaviour field are increasingly attracted by the comparatively good prediction performance of machine learning approaches \citep[cf.][]{HAGENAUER2017273}. 

Of course, the entire purpose of building a forecasting model is not to replicate choices that have already been made, but to make predictions for new settings. Despite the growing interest in machine learning approaches, the dominant, although not necessarily evidenced-based, view is that forecasting models with a solid behavioural underpinning, such as Random Utility Maximisation (RUM) based discrete choice models, are better equipped to predict behaviour under new settings than are models with a weak or no behaviour underpinning, such as e.g. machine learning models \citep[cf.][]{VANCRANENBURGH2022100340}.

However, what is currently less well understood is how to value and incorporate model performance, i.e. the model fit on the empirical data, in the process of developing forecasting models. Performance on empirical (estimation) data is often taken as an indication of a model’s ability to generalise, despite evidence that more flexible models that achieve better in-sample fit do not necessarily lead to improved forecasts \citep[see e.g.][]{Fox_Daly_Hess_Miller_2014,hess2025flexibilityforesightpredictivelimitations}. Moreover, when behavioural patterns in out-of-distribution settings differ from those observed in the estimation data, a good in-sample fit alone may not translate into strong out-of-distribution performance. In such cases, models with a stronger behavioural underpinning may be better suited to capturing underlying decision mechanisms and could therefore generalise more robustly to new scenarios.

The contributions of this paper are twofold. First, we aim to shed light on the prediction performance of forecasting models with varying levels of behavioural underpinning as a function of the ``distance'' between the training data and forecasting scenarios for models. Second, we develop an approach to improve prediction performance through model averaging, where different weights are assigned to different models depending on this ``distance". As highlighted above, there are many ways in which prediction settings could go outside the (multi-dimensional attribute) distribution covered by the estimation data. The core focus of our work is on values of a single continuous attribute,  specifically trip distance. This is not only a natural way of defining the context of a trip, but other key variables such as travel times and cost are correlated with distance, meaning that these variables are also likely to go out-of-distribution for trips that are longer or shorter than those included in the estimation data. There are also policy-relevant motivations for using distance as an example. In particular, forecasting behaviour beyond the range of observed trip distances is of interest in settings where long-distance trips are underrepresented in estimation data. While we use trip distance as the primary variable, the proposed approach is conceptually general and can be applied to other attributes that extend beyond the range observed in the estimation data. It can also be generalised to multi-attribute out-of-distribution metrics. While we focus on trip distance alone in the main body of the paper, such an extension is presented in Appendix \ref{sec: Appendix B}. 

We employ a neural network for the model averaging. This allows learning potential nonlinear relations between distance and model weights. Using the resulting model averaging structure, we aim not only to develop a flexible tool for combining models but also to gain rule-of-thumb insights for the conditions under which specific sorts of models perform best in terms of out-of-distribution forecasting.

The remainder of this paper is organised as follows. Section \ref{sec:methodology} presents our proposed model averaging approach. This is followed by the two case studies in Section \ref{sec:case_studies}, and conclusions in Section \ref{sec:conclusions}.

\section{Methodology}\label{sec:methodology}

As outlined in the introduction, the methodological novelty of the present paper is the use of model averaging for out-of-distribution prediction. In this section, we first discuss the general concept of model averaging before looking into the adaptation required for out-of-distribution prediction.

\subsection{Generic model averaging approach}

Model averaging \citep[see e.g.][]{HANCOCK2020429} can be viewed as a form of stacking, first introduced in the seminal work of Wolpert \citep{wolpert1992stacked}, which established the conceptual basis for ensemble methods combining multiple predictive models. Specifically, we first estimate $M$ different models on the data, with model $m$ giving a likelihood $L_{n,t,m}\left(\Omega_m\right)$ for the choice observed by decision maker $n$ in choice situation $t$, using a vector of estimated model parameters $\Omega_m$. Let us further define $L_{n,m}\left(\Omega_m\right)$ to be the likelihood for the entire sequence of $T_n$ choices observed for person $n$ according to model $m$. 

Model averaging now combines the insights from these different models into one overall structure. The specification of the model averaging log-likelihood function depends on whether an analyst wishes the weights for individual models to be defined at the person level (i.e. $\pi_{n,m}$ for model $m$ for person $n$), or the observation level (i.e. $\pi_{n,t,m}$ for model $m$ for person $n$ in choice situation $t$). With the former, we would have:
\begin{equation}\label{eq:MA_likelihood_person}
LL\left(\Theta\right)=\sum_{n=1}^N log\sum_{m=1}^M\pi_{n,m}L_{n,m}\left(\hat{\Omega}_m\right),	
\end{equation}
while, with the latter, we would have:
\begin{equation}\label{eq:MA_likelihood_obs}
LL\left(\Theta\right)=\sum_{n=1}^N\sum_{t=1}^{T_n} log\sum_{m=1}^M\pi_{n,t,m}L_{n,t,m}\left(\hat{\Omega}_m\right).
\end{equation}
with $T_n$ choices per person. In this notation, $\Theta$ is a vector of parameters for the model averaging model, while $\hat{\Omega}_m$ are the maximum likelihood estimates (MLE) for the vector of parameters $\Omega_m$ for model $m$.

The weights are parameterised as being a function of the person ($z_n$) and observation ($z_{n,t}$) characteristics, with parameters $\gamma_m$ estimated to capture the influence of these characteristics. In particular, we would have:
\begin{equation}\label{eq:pi_person}
\pi_{n,m}\left(\Theta\right)=\frac{e^{\gamma_m'z_n}}{\sum_{l=1}^Me^{\gamma_l'z_n}},
\end{equation}
or
\begin{equation}\label{eq:pi_obs}
\pi_{n,t,m}\left(\Theta\right)=\frac{e^{\gamma_m'z_{n,t}}}{\sum_{l=1}^Me^{\gamma_l'z_{n,t}}},
\end{equation}
where an appropriate normalisation is needed for $\gamma$, typically setting $\gamma_m=0$ for one model $m$.

A reader familiar with latent class models will realise that what is described above is, in essence, a sequential latent class structure. The sequential nature of the process involves first estimating the parameters for the $M$ individual models before then keeping those parameters fixed (at $\hat{\Omega}_m$ for model $m$) and estimating weights for each model, as in Equation \ref{eq:MA_likelihood_person} and \ref{eq:MA_likelihood_obs}. In maximising the model averaging likelihood, an analyst thus, in fact, does not require the model parameters for the individual models, only the probabilities of each observed choice.

In applying the model to predict choices out-of-sample, an analyst then again requires the predicted choice probabilities from the individual models first. Let us assume that $P_{i,n,f,m}\left(\hat{\Omega}_m\right)$ is the probability of person $n$ choosing alternative $i$ in some forecast scenario $f$, according to model $m$, again calculated at the MLEs. Let us further assume that we have worked with model averaging weights at the observation level, i.e. maximising Equation \ref{eq:MA_likelihood_obs} to obtain MLEs for the model weights at the observation level. We can then use these parameters to calculate the model averaging weights for the forecast scenario, i.e. $\pi_{n,f,m}\left(\hat{\Theta}\right)$. The model averaging probability for choosing alternative $i$ in this new setting would then be given by:
\begin{equation}\label{eq:MA_pred}
\hat{P}_{i,n,f,MA}=\sum_{m=1}^M\hat{\pi}_{n,f,m}\left(\hat{\Theta}\right)P_{i,n,f,m}\left(\hat{\Omega}_m\right).
\end{equation}

\subsection{Model averaging for out-of-distribution prediction}

Model averaging typically involves estimating each model on the full dataset and then computing weights based on how well each model fits for each of the data points, estimating such weights at the person or observation level. These weights can depend on decision-maker characteristics such as age, gender, and income, as well as trip characteristics such as purpose and trip distance. In our proposed approach, the weights instead stem from a characterisation of how far an observation is from the remainder of the data, i.e. a measure of being out-of-distribution. Different weights will then be given to different models depending on this measure, with for example some models being given increased weight when predicting choices for less ``typical'' observations. It should be noted that this definition of out-of-distribution prediction may favour models with relatively high bias and low variance, as such models often extrapolate more smoothly beyond the observed range \citep{bishop2006pattern}.

In the present paper, as outlined in the introduction, we focus specifically on trip distance as the characteristic determining how ``typical'' an observation is. This is partly a pragmatic choice, reflecting the proof-of-concept nature of our work. It allows for a straightforward definition of in-distribution and out-of-distribution settings through a simple data split, facilitating a clear assessment of the proposed model averaging approach. Trip distance also has several appealing characteristics in this context. First, it is typically not included directly as an explanatory variable in the mode choice models, which instead rely on closely correlated variables such as time and cost. This in itself is not a requirement though - there are no a priori reasons why a variable used in the individual models could not be used as a variable in the model averaging weight function. Second, distance provides a parsimonious way of capturing variation across trips, as it is correlated with multiple underlying attributes. This allows the model averaging structure to learn how model performance varies with trip context using a single, interpretable variable, rather than relying on a higher-dimensional set of covariates. The exact same approach can be used with a multivariate measure, as discussed in Appendix \ref{sec: Appendix B}.

Training a model averaging structure where trip distance, say $d_{n,t}$ for person $n$ and trip $t$, is used as a key characteristic in Equation \ref{eq:pi_obs} would allow the model to learn how different sub-models are better suited for predicting trips of specific distances within the bounds of the estimation data. However, if the individual sub-models have been estimated on the entire data, then model averaging by definition cannot gain any insights about how the performance might differ across models out-of-distribution, and thus what weights should be given to each model outside the range of distances used in estimation. As a result, the weights would simply be an extrapolation based on the estimates for $\gamma$ obtained on the training data. For example, imagine that our model would find that the weight assigned to model $m$ increases with distance. If making predictions out-of-distribution, the model would follow the same trend, and for trips with distances below the lower bound of the estimation data, the weight assigned to model $m$ would decrease, while for trips with distances above the upper bound of the estimation data, the weight assigned to model $m$ would increase.

In the present paper, we propose a different approach that enables us to examine the weight assigned to individual models in out-of-distribution predictions. Again, using distance as an example, let us assume that for the available estimation data, we have minimum and maximum distances $d_{min}$ and $d_{max}$, i.e. $D_{data}=\left[d_{min},d_{max}\right]$. We would then define a narrower interval $D_{train,sub}=\left[d_{a},d_{b}\right]$, where $d_{min}<d_{a}<d_{b}<d_{max}$. In the estimation of the individual sub-models, we then use only a subset of the range of the available data, specifically those observations where $d_{n,t}\in D_{train,sub}$. The model averaging structure, on the other hand, will be estimated on the entire data, i.e. still maximising e.g. Equation \ref{eq:MA_likelihood_obs}. This means that the probabilities for the individual models will be calculated for all trips in the data, but using parameters $\hat{\Omega}_m$ obtained from optimisation on only a part of the data (where $d_{n,t}\in D_{train,sub}$). Model averaging learns the role of trip characteristics in determining which model should obtain more weight for a specific trip, in our case focusing on journey distance, but with easy extension to other variables, which would be grouped together in the vector $z_{n,t}$ in Equation \ref{eq:pi_obs}. By splitting the data in the way described above, we can now also define new variables to be used in Equation \ref{eq:pi_obs} that help characterise whether a given observation is within $D_{train,sub}$, and if not, how far outside the training data it is located, allowing the model averaging process to gain insights into extrapolation beyond the estimation space of the original sub-models. Indeed, the above approach directly allows the model averaging procedure to learn how individual sub-models perform outside the range of distances on which they have been trained and how the weight given to each should, as a result, change the further we move away from that interval. This, in turn, then also means that after estimating the model averaging structure, predictions can be made outside the range $D_{data}$, which is of key interest in travel demand forecasting. The model averaging procedure will have \emph{learned} how the weight assigned to different models changes outside the estimation data and can then apply that knowledge in out-of-distribution prediction. For example, the model averaging process might determine that models with a stronger behavioural foundation suffer a smaller drop in performance outside the estimation data and will thus gain weight in out-of-distribution prediction. 

It should be clear that an analyst will need to make important trade-offs in this process. By increasing the gaps $d_{a}-d_{min}$ and $d_{max}-d_{b}$, the analyst will increase the ability of the model averaging structure to learn about out-of-distribution weights. But this will come at the cost of reducing the width of $D_{train,sub}$ and the size of the sample used for training the individual sub-models, reducing the ability to obtain robust estimates of the influence of level-of-service variables, for example.

Hitherto, model-averaging studies have primarily relied on logistic regression models, as shown in \eqref{eq:pi_obs}, often with linear specifications. Because of this, such models are ill-equipped to capture complex relations between explanatory variables used for the model averaging and the likelihood of the model $\pi_{n,m}$. However, when it comes to out-of-distribution prediction using model-averaging, nonlinear effects, e.g., of distance, can be expected. Therefore, in this study, we propose using an MLP for the model averaging, which is much more flexible and can learn nonlinearities and interactions between explanatory variables from the data. More specifically, the MLP is trained to learn the function $g()$, which takes the model probabilities, $P_{n,m}$ and explanatory variables $Z_{n}$ to produce the model weight: \( {\pi}_{i,n,m} = g(\omega \mid P_{n,m}, z_{n}) \). The flexibility of neural networks is also likely to be especially beneficial when using multiple characteristics in the measure of ``typicality''.

\section{Case studies} \label{sec:case_studies}

In this paper, we illustrate the performance of the proposed approach in two case studies. In what follows, we first describe the two datasets, before talking about how the data was divided for the purpose of the present paper. We then talk about the individual sub-models and the specification and estimation of the model averaging structure. We finally look at the results.

\subsection{Data} 	

We use two different revealed preference datasets in this study, both focusing on mode choice.

\subsubsection{DECISIONS data}

The first dataset comes from a large-scale survey conducted as part of the DECISIONS project carried out by the Choice Modelling Centre at the University of Leeds \citep{calastri2020we}. We specifically use the GPS tracking component of this survey, where the data used for this work corresponds to the observed mode choice behaviour. Respondents were tracked over a period of two weeks, with most trips taking place in Yorkshire. 

After extensive data cleaning and data enrichment \citep{tsoleridis2022deriving}, 10,990 trips made by 415 individuals remained. For each trip, individuals travelled by one of six modes: car (48\%), bus (14\%), rail (5\%), taxi (3\%), cycling (4\%) or walking (26\%). The data covers a wide range of trip distances, going from just under 100 metres to just under $106$ kilometres, with a mean distance of $8.2$ kilometres and a median distance of $4.5$ kilometres. It thus covers both urban trips as well as inter-city trips, for example between Leeds and York. 

The attributes of the alternatives used in the models for the present paper include in-vehicle travel time, out-of-vehicle travel time, and travel cost.

\subsubsection{LPMC data}

The second dataset is the London mode choice data compiled by \citet{jsmic.17.00018}, referred to hereafter as LPMC (London passenger mode choice). 

We use a sample of $80,943$ trips made by $31,921$ individuals. This dataset contains four alternatives: walking (18\%), cycling (3\%), public transport  (35\%, grouping together bus and rail) and driving (44\%, grouping together private car and taxi rides). The data again covers a wide range of trip distances, going from just under 100 metres to just over $40$ kilometres, with a mean distance of $4.6$ kilometres and a median distance of $2.8$ kilometres. This dataset covers all of London, including short-distance trips as well as journeys from one side of London to the other.

Attributes of the alternatives used in the models include in-vehicle travel time, out-of-vehicle travel time, interchanges, and travel cost, along with two sociodemographic variables, namely car ownership and driving licence status.
 
\subsection{Estimation and test data}\label{sec:data_splitting}

As outlined in the earlier discussions, our approach to model averaging relies on the idea of excluding part of the distance distribution in the training of the individual sub-models, by setting $d_{a}>d_{min}$ and $d_{b}<d_{max}$. For the purpose of testing the performance of the model averaging approach, we go one step further in our empirical work, by excluding very short and very long trips from the training of the model averaging structure as well, thus allowing us to validate the out-of-distribution performance.

Formally, we have that the range of distances in the data is given by $D_{data}=\left[d_{min},d_{max}\right]$. The range used in training the individual sub-models is given by $D_{train,sub}=\left[d_{a,sub},d_{b,sub}\right]$. Finally, the model averaging models are estimated on $D_{train,MA}=\left[d_{a,MA},d_{b,MA}\right]$. Overall, the relationship holds that $d_{min}<d_{a,MA}<d_{a,sub}<d_{b,sub}<d_{b,MA}<d_{max}$. For the purposes of our analysis, we set $d_{a,MA}$ and $d_{b,MA}$ to be at the 10$^{th}$ and 90$^{th}$ percentiles of the sample distance distribution $D_{data}$, while $d_{a,sub}$ and $d_{b,sub}$ are set at the 20$^{th}$ and 80$^{th}$ percentiles of the sample distance distribution $D_{data}$.

As outlined above, the shortest 10\% and longest 10\% of trips were retained for out-of-distribution testing. In both datasets, this implies that the out-of-distribution portion corresponds to the tails of the distance distribution. The upper tail is particularly extensive, covering trips between 19.52 km and 105.9 km in the DECISIONS data, and between 11.22 km and 40.1 km in the LPMC data. This represents over 80\% of the distance range in the DECISIONS data and over 70\% in the LPMC data. This creates a meaningful out-of-distribution setting, with inter-city trips in the DECISIONS data and cross-London trips in the LPMC data appearing exclusively in the test set, meaning that the contexts are structurally different from the estimation data.

\begin{figure}[h]
\centering
\includegraphics[width=15cm]{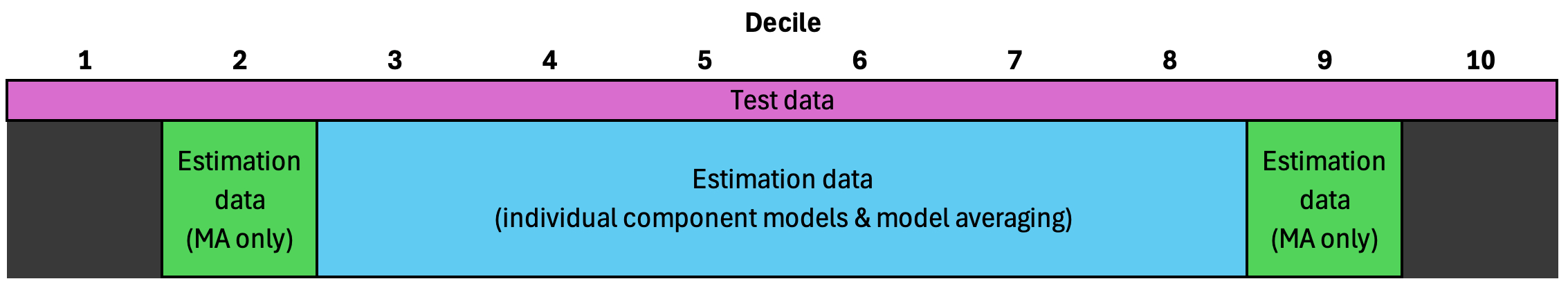}
\caption{Estimation and test data}	
\label{fig:data_split}
\end{figure}

In addition to these divisions, we further retain $20\%$ of any trips, independent of their distance, for out-of-sample test. The splitting of the sample into estimation, test (and additionally validation data for MLP and XGB) was done at the person/household level, to avoid data leakage. In other words, each person only contributed data to either the estimation data or the test data. For those individuals who contributed to the estimation data, their observations in distance segments 1 and 10 were thus not used in our analysis. The resulting division of the data is explained in Figure \ref{fig:data_split}, with the sample sizes shown Table \ref{tab:data_split}. In practical work, analysts would be unlikely to not include all deciles (intervals of 10\% width) in the MA estimation work; we made these decisions with a view to validating the model idea. Similarly, a real-world application would not set aside the test data, and thus also avoid deleting observations in the estimation data for trips that are in segments 1 and 10 (see the earlier point about data leakage).

\begin{table}[h]
  \vspace{0.5cm}
  \centering
  \caption{Estimation and test data sample sizes for case studies}
  \resizebox{\textwidth}{!}{%
    \begin{tabular}{rcccccccc}
          \toprule
          & \multicolumn{4}{c}{\textbf{DECISIONS data}} & \multicolumn{4}{c}{\textbf{LPMC data}} \\
          & \multirow{2}{*}{\textbf{Distance (km)}} & \multicolumn{2}{c}{\textbf{Estimation}} & \multirow{2}{*}{\textbf{test}} & \multirow{2}{*}{\textbf{Distance (km)}} & \multicolumn{2}{c}{\textbf{Estimation}} & \multirow{2}{*}{\textbf{test}} \\
          &       & \textbf{Individual} & \textbf{MA} &       &       & \textbf{Individual} & \textbf{MA} &  \\
          \midrule
    \textbf{out of distribution for MA (below, i.e. $d_{min}$ to $d_{a,MA}$)}\unboldmath{} & <1.33 & 0     & 0     & 223   & <0.75 & 0     & 0     & 1,581 \\
    \textbf{out-of-distribution for sub-models (below, i.e. $d_{a,MA}$ to $d_{a,sub}$)}\unboldmath{} & 1.33 to 1.95 & 0     & 836   & 261   & 0.75 to 1.11 & 0     & 6,446 & 1,642 \\
    \textbf{within distribution for MA and sub-models (i.e. $d_{a,sub}$ to $d_{b,sub}$)} & 1.95 to 11.95 & \multicolumn{2}{c}{5,272} & 1,322 & 1.11 to 7.41 & \multicolumn{2}{c}{38,761} & 9,790 \\
    \textbf{out-of-distribution for sub-models (above, i.e. $d_{b,sub}$ to $d_{b,MA}$)}\unboldmath{} & 11.95 to 19.52 & 0     & 946   & 153   & 7.41 to 11.22 & 0     & 6,458 & 1,646 \\
    \textbf{out of distribution for MA (above, i.e. $d_{b,MA}$ to $d_{max}$)}\unboldmath{} & 19.52 to 105.9 & 0     & 0     & 154   & 11.22 to 40.1 & 0     & 0     & 1,630 \\
   \midrule
   \textbf{total} &       & \textbf{5,272} & \textbf{1,782} & \textbf{2,113} &       & \textbf{38,761} & \textbf{12,904} & \textbf{16,289} \\
    \bottomrule
    \end{tabular}%
  }
  \label{tab:data_split}
\end{table}

\subsection{Model types} \label{subsec:model_types}

The following individual models are used in our analysis, combining structures from choice modelling, mathematical psychology and machine learning. Many other model types could have been considered\footnote{A notable omission in our list is mixture models such as mixed logit or latent class. Any advantages that such models have in estimation largely/completely disappear in prediction as the random heterogeneity needs to be averaged out. The only exception comes with the use of posterior distributions, which are not applicable in the context of out-of-sample prediction. For a discussion of this issue, see \citet{hess2025flexibilityforesightpredictivelimitations}.}, where we limited ourselves to this set with a view to demonstrating the potential benefit of our proposed approach. 

\begin{description}
	\item[Nested Logit (NL):] Our NL models \citep[cf.][]{154} use a full set (i.e. $J-1$) of alternative specific constants (ASC). For continuous attributes, we use a linear plus logarithmic specification, i.e. $V=...+\beta_x x+\beta_{log-x}log(x)$ to capture non-linearity in sensitivities. For the DECISIONS data, we use three nests, grouping car together with taxi, bus with train, and walking with cycling. For the LPMC data, we initially estimated a NL model using two nests, grouping car together with bus, and walking together with cycling, but this structure ended up collapsing to a Multinomial Logit (MNL) model.
	\item[Decision field theory (DFT):] DFT is a dynamic, stochastic model, introduced by \citet{busemeyer1993decision}. The key idea of the DFT model is that the preferences for different alternatives update over time whilst the decision-maker considers the different alternatives and their attributes. We use the implementation of \citet{hancock2018accumulation}, with the valence functions following the same specifications as the utility in NL/MNL.
	\item[MultiLayer Perceptron (MLP):] This model comprises an input layer with input nodes, one or more hidden layers with hidden nodes, and an output layer with output nodes. In this model, signals propagate forward through the links connecting the nodes. The links have numeric weights $w$, which are learned from the data. At each link, the weights are multiplied by the input value from the previous nodes. At the node, the sum of the inputs is taken, and an activation function is applied. The outcome is propagated to the nodes in the next layer. We use \textit{tanh} activation functions in all hidden layers, except the final one, where a linear layer is used. In the output layer, a Softmax function (i.e. a logit) is applied to produce choice probabilities for each alternative. To ensure consistency with the behavioural models and enhance predictive performance, availability conditions are explicitly incorporated into the network (thus, treated in the same way as in traditional choice models). 
	\item[XGBoost (XGB):] The XGB model comprises a series of sequentially applied decision trees. A decision tree is a sequence of simple IF-THEN rules, optimised to classify data accurately. In the XGB model, each decision tree in the series `corrects' the mispredictions of the models before it. This process is referred to as `boosting'. The term ‘gradient’ refers to the fact that each successive tree is fitted using the gradient of the loss function—effectively via gradient descent. Unlike standard Gradient Boosting Machines (GBM), XGBoost includes further optimisations, such as second-order gradient information and advanced regularisation, to improve speed and mitigate overfitting. To ensure consistency with the behavioural models and to better handle stronger regularisation settings, availability conditions are enforced using \textit{base margin}: we set margins to 0 for available alternatives and a large negative value for unavailable ones, ensuring that the model assigns zero probability to non-available alternatives. 
 \end{description}

\noindent The NL and DFT models were coded and estimated using Apollo \citep[cf.][]{hess_palma_apollo}, while the MLP and XGB models were implemented in Python. The MLP was implemented in PyTorch \citep{NEURIPS2019_9015} using a custom  to handle the availability conditions, while the XGBoost model was implemented directly with the \textit{xgboost} package \citep{Chen:2016:XST:2939672.2939785}. 

For the two data-driven models, a hyperparameter grid search was conducted to optimise performance. We used a 5-fold grouped cross-validation procedure to account for the panel structure of the data (see Table \ref{tab:hyperparamspaces}). 

For MLP, the number of hidden layers and nodes per layer influence the model’s capacity to learn complex patterns, while the batch size affects the stability and efficiency of the gradient updates. The learning rate controls how quickly the model updates its weights during training, while the L2 regularisation helps prevent overfitting by penalising large weight values. Since the best-performing hyperparameters were highly similar across both datasets, we opted to use a single configuration. The best-performing configuration found was two hidden layers with 30 nodes each, a learning rate of 0.001, a batch size of 250 and an L2 regularisation strength of 0. 

An important point to highlight is variability in results. Each training run of an MLP can yield slightly different outcomes. An analyst then needs to make a decision on how many training runs to use, and which of the results to retain. We trained the model 100 times using the configuration described above. If our focus at this stage was purely on prediction performance, we would concentrate on the best fitting models out of these runs. However, our objective here is also to consider diversity across model instances, capturing variability arising from different
local optima and thereby promoting ensemble diversity. We thus averaged the predictions across these 100 runs to account for this variability.

The hyperparameter ranges for the XGB model were informed by the results of \citet{hillel2021new}. The hyperparameter search included the maximum tree depth, the subsample fraction, the fraction of features considered per split, regularisation strengths (L1 and L2), and the maximum delta step. Furthermore, we set the learning rate to 0.1 and used the default values for all other hyperparameters. During training, early stopping was used to control the number of boosting rounds (with a maximum set to 3000). Thus, training was stopped if adding more trees did not further improve performance (on the validation set) after 10 boosting rounds. A number of hyperparameters are considered:
\begin{itemize}
\item the maximum tree depth controls the complexity of each tree, with deeper trees allowing more detailed partitions at the risk of overfitting; 
\item the subsample fraction regulates the proportion of training samples used for each boosting round, and the feature subsampling parameter (colsample by tree) increases model robustness by introducing randomness at the feature level;
\item L1 and L2 regularisation terms penalise overly large weights and help stabilise the model under strong regularisation; and 
\item the maximum delta step restricts the weight update size, improving convergence stability in cases of unbalanced data. 
\end{itemize}
The optimal hyperparameters for LPMC were: \textit{Max. tree depth} = 4, \textit{Subsample fraction} = 0.8, \textit{Colsample by tree} = 0.8, \textit{$L_1$ reg.} = 10, \textit{$L_2$ reg.} = 1, and \textit{Max. delta set} = 1. 
The optimal hyperparameters for DECISIONS were: \textit{Max. tree depth} = 2, \textit{Subsample fraction} = 0.4, \textit{Colsample by tree} = 0.6, \textit{$L_1$ reg.} = 1, \textit{$L_2$ reg.} = 20, and \textit{Max. delta set} = 2.

As with the MLP model, we repeated the XGB training 100 times using the optimal configuration and averaged the predictions to account for variability in the model training. 

\begin{table}[h]
    \centering
    \vspace{0.5cm}
    \caption{Hyperparameter spaces considered for the MLP and XGBoost models}
    \label{tab:hyperparamspaces}
    \begin{tabular}{p{7.0cm} p{7.0cm}}
        \hline
        \textbf{MLP} & \textbf{XGBoost} \\
        \hline
        \begin{tabular}[t]{@{}l l@{}}
        Hidden layers:     & $\{(10,5), (10,10),$ \\
                          & $\;(20,20), (30,30)\}$ \\
        Batch size:        & $\{250, 500, 1000\}$ \\
        Learning rate:     & $\{0.01, 0.001, 0.0001\}$ \\
        $L_2$ regularisation: & $\{0, 0.1, 0.5, 1\}$ \\
        \end{tabular}
        &
        \begin{tabular}[t]{@{}l l@{}}
        Max. tree depth:      & $\{2, 3, 4, 5\}$ \\
        Subsample fraction:   & $\{0.4, 0.6, 0.8\}$ \\
        Colsample by tree:    & $\{0.4, 0.6, 0.8\}$ \\
        $L_1$ regularisation:    & $\{1, 5, 10, 20\}$ \\
        $L_2$ regularisation:    & $\{1, 5, 10, 20\}$ \\
        Max. delta step:      & $\{0, 1, 2, 4\}$ \\
        \end{tabular} \\
        \hline
    \end{tabular}
\end{table}

\subsection{Model averaging: specification and estimation}\label{sec:MA}

Similar to the MLP models used for training on the datasets (see Section~\ref{subsec:model_types}), the MLP used for model averaging was implemented in PyTorch \cite{NEURIPS2019_9015}. This implementation is required because the loss function (see Equation~\eqref{eq:MA_likelihood_obs}) involves a custom specification that cannot be readily implemented in, for example, Scikit-learn. As explanatory features, we included three distance-based variables: the raw trip distance, its logarithmic transformation, and its squared term. While in principle, the linear distance alone should suffice, we found that including the logarithmic and squared transformations resulted in a small but consistent improvement in predictive performance. The fact that we included multiple variants of the same variable in our MA structure already shows the simplicity of expanding the approach for using multiple different variables.

\begin{figure}[t!]
    \centering
    \includegraphics[width=0.8\linewidth]{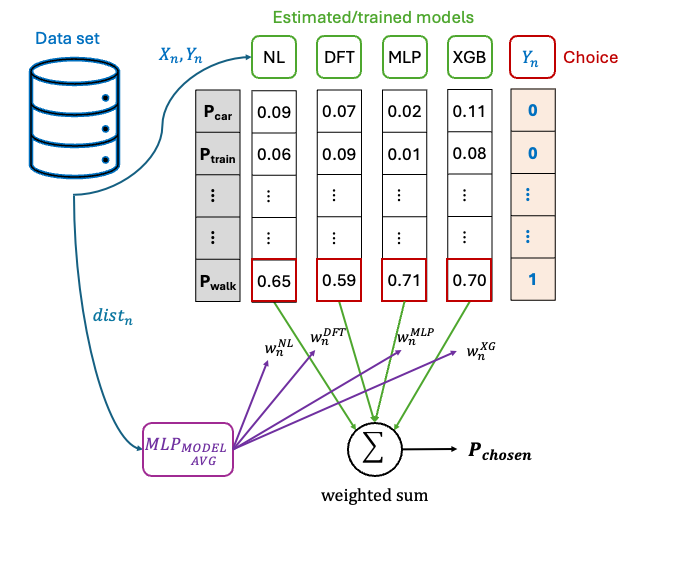}
    \caption{Model averaging structure}
    \label{fig:MA_structure}
\end{figure}

Figure \ref{fig:MA_structure} shows the structure for our model averaging approach. As shown in the figure, the model averaging MLP only takes the trip distance features as input to predict the weights given to each sub-model. The model-averaging probability is computed as a weighted sum, combining the predictions for the trip by each sub-model and the weights assigned to each sub-model by the model-averaging MLP.

Hyperparameters were tuned using a grid search combined with five-fold cross-validation, where folds were defined at the individual (\texttt{household\_id}) level to avoid leakage across observations from the same individual. The following parameter spaces were considered: batch size (50, 100, 250), one hidden layer with 10, 20, or 30 nodes, learning rate (0.01, 0.001, 0.0001), and L2 regularisation strength (0, 0.01, 0.05). 
For the DECISIONS dataset, the best-performing model has a hidden layer with 10 nodes, a batch size of 50, a learning rate of 0.01, and an L2 regularisation strength of 0. 
For the LPMC dataset, the best-performing model has a hidden layer with 10 nodes, a batch size of 250, a learning rate of 0.01, and an L2 regularisation strength of 0.

Using the optimal hyperparameters, we repeated the training 100 times. In contrast with the training of the individual models, our objective is now high predictive performance. We therefore retained only the top 20 per cent of models\footnote{This selection was based on the log-likelihood. For this, we used the full data, i.e.\ training and validation data, but excluded the test data.}, discarding those that became trapped in poor local maxima. This approach differs slightly from the one used for training the data-driven models, where we also repeated training 100 times but averaged predictions across all models to encourage diversity, while diversity is now captured by the different models entering the ensemble.

It should be noted that the data used to train the meta-model are the same observations used to train the base models (for distance segments 3–8). \cite{hastie2009elements} argue that using the same data for training both the base models and the meta-model may lead the meta-model to assign disproportionately high weights to more complex models, as their in-sample predictions can appear overly optimistic. To mitigate this issue, they propose a stacking procedure in which the meta-model is trained on predictions generated from data not used in training the base models (often obtained via cross-validated predictions)\footnote{To further facilitate comparison with the machine learning literature on stacked generalisation, Table~\ref{tab:terminology_mapping} provides a mapping between the terminology used in this paper and the standard terminology used in that literature.}.

While we recognise the concern about using the same data for both the base models and meta-model, we deliberately refrained from implementing a stacking procedure in the main analysis in order to keep the modelling pipeline tractable. The overall framework already involves a large number of models, data partitions, hyperparameter tunings, and aggregation steps, and introducing a full cross-validated stacking procedure would substantially increase the computational and methodological complexity. Nevertheless, to assess whether this modelling choice affects our results, we conducted an additional analysis on the larger LPMC dataset in which the meta-model is trained on predictions generated from data not used to train the base models. The results of this analysis, reported in Appendix \ref{sec: Appendix A}, indicate that the impact on predictive performance is negligible, suggesting that the simpler procedure used in the main analysis does not materially affect the conclusions, at least in our application.

\begin{table}[h!]
\centering
\caption{Terminology mapping between this study and the stacking literature}
\label{tab:terminology_mapping}
\begin{tabular}{ll}
\hline
\textbf{Current paper} & \textbf{Stacking literature} \\
\hline
Sub-model / individual model & Base learner (level-0 model) \\
Model averaging model & Meta-learner (level-1 model) \\
Model averaging weights & Combination weights / stacking weights \\
Predicted probabilities (per model) & Base learner predictions \\
Ensemble / model averaging approach & Stacked generalisation / stacking \\
Distance-based features & Meta-features \\
\hline
\end{tabular}
\end{table}

\subsection{Empirical results}

In this section, we present the empirical results for the two case studies. We first focus on the performance of the individual sub-models, comparing this across distance segments and also discussing out-of-sample performance. We then look at the results for the model averaging structure. 

\subsubsection{Performance of individual sub-models}

Table \ref{tab:LL_estimation} presents the performance of the different models on the estimation sample, covering distance segments (deciles) 3 to 8, as illustrated in Figure \ref{fig:data_split}. 

We see that for the DECISIONS data, NL obtains a higher log-likelihood (LL) than DFT, while DFT outperforms MNL in the LPMC data. In both studies, the two machine learning approaches obtain much higher LL than NL/MNL and DFT. Proportionally, the difference is slightly larger in the DECISIONS case study than in the LPMC one. In the former, MLP outperforms XGB, while the reverse applies for the latter. No formal statistical tests are carried out to compare the different model structures as these are not applicable to the machine learning structures.

\begin{table}[h]
    \centering
    \vspace{0.5cm}
    \caption{Final log-likelihood of different model structures (covering distance segments 3-8) of the estimation set}
    \label{tab:LL_estimation}
    \begin{tabular}{rcc}
        \toprule
        Model & \multicolumn{1}{c}{DECISIONS} & \multicolumn{1}{c}{LPMC} \\
        \midrule
        NL/MNL  & -2,319.90  & -28,706.81 \\
        DFT & -2,324.74  & -28,670.31 \\
        MLP & -2,082.31  & -27,600.70 \\
        XGB & -2,171.23  & -26,195.36\\
        \bottomrule
    \end{tabular}
    \vspace{0.5cm}
    \end{table}

\begin{sidewaysfigure}[hp!]
\includegraphics[width=10cm]{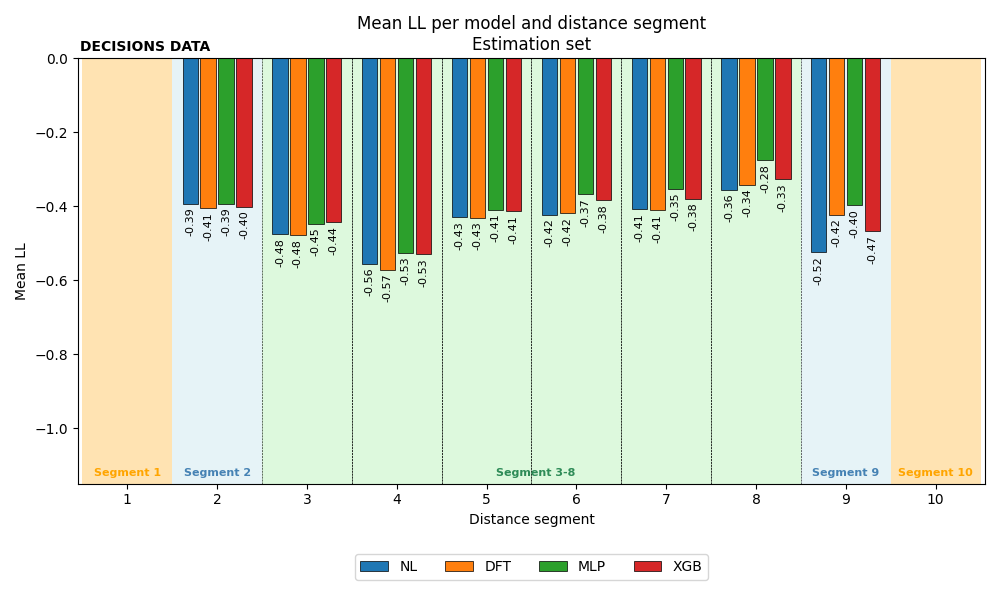}
\includegraphics[width=10cm]{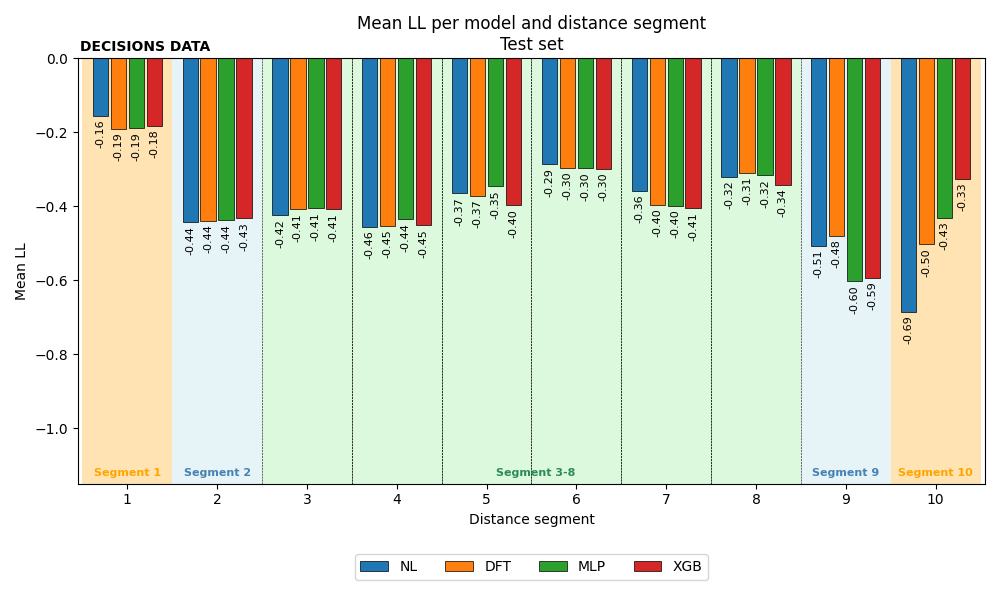}\\
\includegraphics[width=10cm]{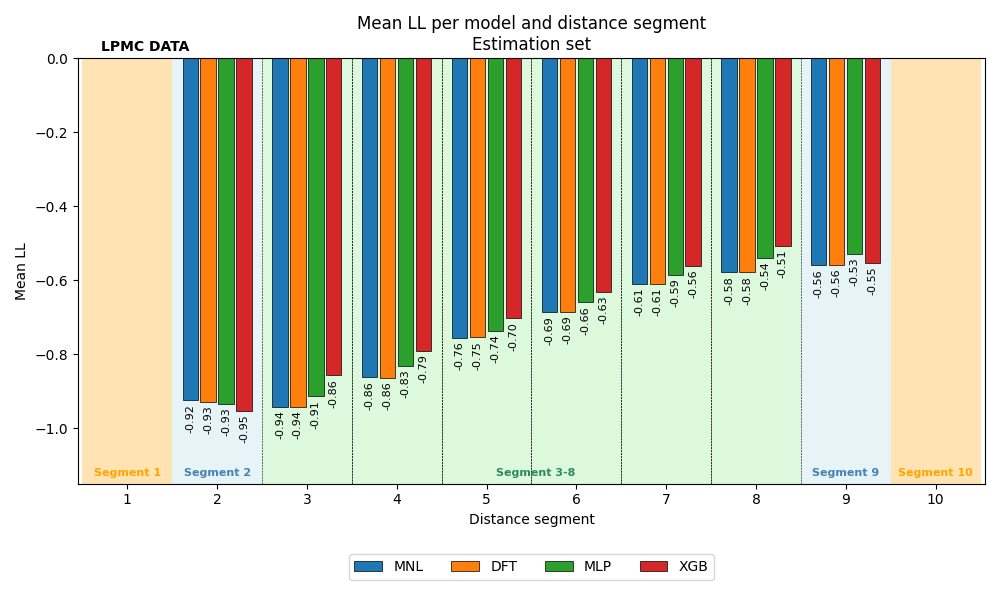}
\includegraphics[width=10cm]{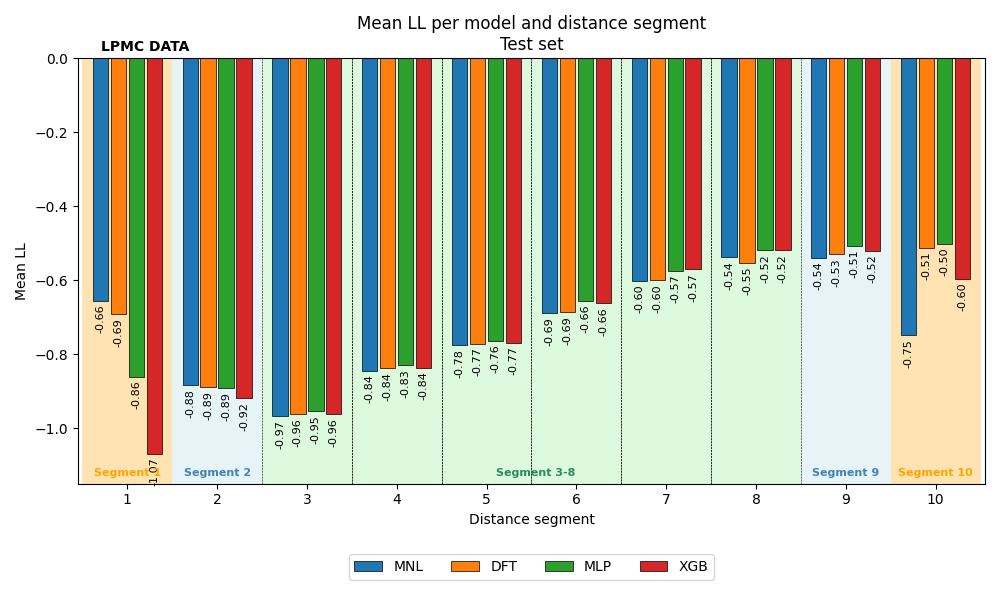}
\caption{Estimation and test performance}	
\label{fig:est_test}
\end{sidewaysfigure}

\noindent As a next step, Figure \ref{fig:est_test} looks at the performance of these models across the different distance segments obtained by splitting the data into deciles of the distance distribution. We do this separately for the estimation and test data. A point to note when studying these results is that for the individual models, only segments 3 to 8 were used in estimating the model parameters, but the probabilities in segments 2 to 9 for the individual models also contributed to the model averaging, as explained in Figure \ref{fig:data_split}. The values shown for segments 2 and 9 in the left two graphs in Figure \ref{fig:est_test} thus in effect also relate to out-of-distribution validation for the individual models.

Starting with the graphs on the left, we see that, across the two datasets, the advantage for MLP and XGB that we noted at the overall level in Table \ref{tab:LL_estimation} applies in each of the 6 distance segments used in estimation (i.e. segments 3-8). Across the distance segments, the same trend as overall (XGB outperforming MLP) is observed in all six segments (3-8) for LPMC. However, for the DECISIONS data, the overall trend (MLP outperforming XGB) stems from the large differences in segments 6-8 only, showing a more diverse pattern. Meanwhile, the overall trend between DFT and NL/MNL is reversed in some cases in both datasets. These findings already show that different models perform differently well across trip distances. Interestingly, while the LPMC results suggest that mode choice is easier to predict for longer trips (across segments 3-8), no clear trend emerges for the DECISIONS data.

We next turn to segments 2 and 9 in the left two graphs, giving us initial insights into out-of-distribution performance, remembering that the values shown here are the predicted probabilities for the chosen alternative, using the model parameters obtained when estimating the models on segments 3 to 8. In the DECISIONS data, we see all models perform very similarly in segment 2, while in segment 9, DFT outperforms XGB. In the LPMC models, NL and DFT perform best in segment 2, while MLP does best in segment 9, but differences are again small.

We next turn to the test results, shown in the right two graphs in Figure \ref{fig:est_test}. We first focus on segments 3-8 with a view to understanding potential overfitting to the estimation data. For the DECISIONS data, we see some evidence of improved out-of-sample (vs within-sample) performance for all models, with an average gain in per observation LL of $0.068$ for NL, $0.065$ for DFT, $0.025$ for MLP, and $0.025$ for XGB. For the LPMC models, there are improvements by $0.008$ for NL and DFT, a change of $<10^{-5}$ for MLP and a drop by $0.04$ for XGB. Overall, these results do not suggest any overfitting, at least within the distribution of distances covered by the estimation data.

The findings for segments 2 and 9 on the test data largely confirm those we saw in the two graphs on the left in Figure \ref{fig:est_test}, except for MLP and XGB performing more poorly in the test data for segment 9 for the DECISIONS data. We finally turn to segments 1 and 10. From a behavioural perspective, we see that for the DECISIONS data, mode choice on very short trips is much easier to predict. For short trips in the LPMC dataset, there is a clear advantage for the behavioural models. For the longest trips (segment 10), the trend is less clear. In the DECISIONS data, MLP and XGB perform comparatively well. In the LPMC data, both data-driven models again do comparatively well, while MNL experiences a substantial decline in performance.

\subsubsection{Model averaging results}

In model averaging, data covering segments 2-9 were used, but with model parameters for the individual models estimated on segments 3-8 only (cf. discussion around Figure \ref{fig:data_split}). Table \ref{tab:LL_MA_estimation} shows the performance (in terms of LL on the estimation set) of the four individual models, alongside that of the model averaging (MA) structure. In both datasets, the MA structure clearly obtains a higher LL than any of the four individual models. It is noteworthy that this finding is robust even to variations in the performance of the individual models. In other words, the MA structure attains a similarly high log-likelihood even when one or more of the constituent models perform less well. This echoes the well-known principle in machine learning that an ensemble of weak learners can together form a strong learner. From a practical perspective, this suggests that analysts need not exhaustively optimise each individual model when developing a model-averaging framework.

\begin{table}[h]
    \centering
    \vspace{0.5cm}
    \caption{Log-Likelihood for model averaging: individual components and overall structure (covering distance segments 2-9), with gains by MA over individual models shown in brackets}
    \label{tab:LL_MA_estimation}
    \begin{tabular}{rcc}
        \toprule
        Model & \multicolumn{1}{c}{DECISIONS} & \multicolumn{1}{c}{LPMC} \\
        \midrule
    NL/MNL & \multicolumn{1}{l}{-3,145.33 (12.48\%)} & \multicolumn{1}{l}{-38,282.79 (7.25\%)} \\
    DFT    & \multicolumn{1}{l}{-3,063.46 (10.14\%)}  & \multicolumn{1}{l}{-38,263.11 (7.2\%)} \\
    MLP    & \multicolumn{1}{l}{-2,787.34 (1.24\%)}  & \multicolumn{1}{l}{-37,038.34 (4.14\%)} \\
    XGB    & \multicolumn{1}{l}{-2,949.47 (6.67\%)}  & \multicolumn{1}{l}{-35,920.99 (1.15\%)} \\
    \textbf{MA} & \textbf{-2,752.68} & \textbf{-35,506.69} \\
        \bottomrule
    \end{tabular}
    \vspace{0.5cm}
\end{table}

\begin{figure}[h]
\includegraphics[width=15cm]{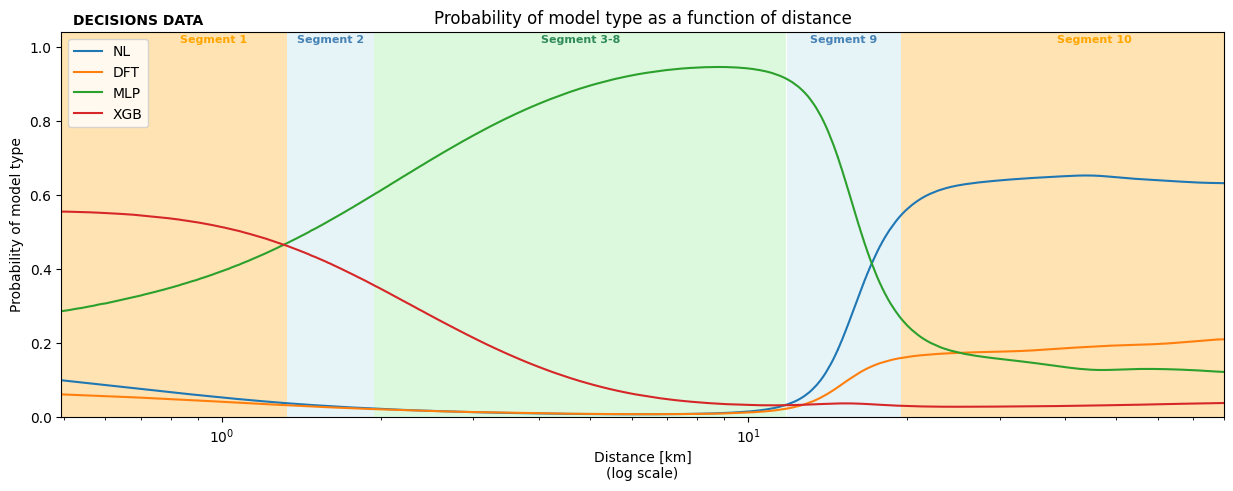}\\
\includegraphics[width=15cm]{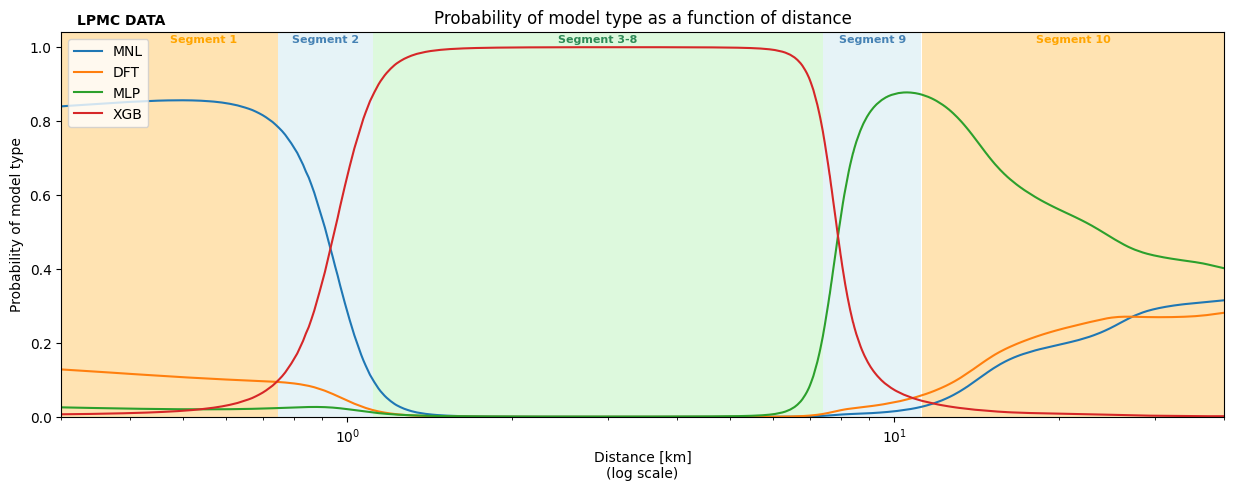}
\caption{Model averaging: weights given to individual models}	
\label{fig:MA}
\end{figure}

\noindent The inclusion of segments 2 and 9 allows the model averaging structure to learn about out-of-distribution performance and hence guide the process of linking the weights given to individual models to the difference between the distance for that trip and the data used in estimation (i.e. segments 3 to 8). As explained in Section \ref{sec:MA}, a flexible neural network is used to estimate this relationship, with no a priori assumptions about the functional form. The resulting relationship is illustrated in Figure \ref{fig:MA}.

For both datasets, we see that the MA process gives almost the entire weight to MLP and XGB for any trips with a distance contained within $D_{train,sub}$, i.e. distances on which the individual sub-models were estimated (segments 3-8). In the DECISIONS data, the weight of XGB increases on shorter trips within that part of the distance distribution compared to long trips, while for longer trips, almost the entire weight goes to MLP for DECISIONS. For LPMC, all the weight goes to XGB. These findings are in line with the pattern of prediction performance obtained for individual models, as shown on the left-hand side in Figure \ref{fig:est_test}. 

Once we go outside the $D_{train,sub}$ interval, the pattern changes. In both datasets, we observe that for trips with distances below $d_{a,sub}$ (i.e. segments 1\&2), the weight for behavioural models increases. In the DECISIONS data, only a small share goes to NL and DFT. On the other hand, in LPMC, MNL obtains over 80\% of the weight. For longer trips, i.e. above $d_{b,sub}$, NL gets the vast majority of the weight in the DECISIONS data, with DFT getting more than MLP or XGB. For LPMC, MLP retains a large weight, but this decreases while the weight for MNL and DFT increases with distance.

The final step in our analysis of results concerns the performance of the model averaging structure on the test data. Table \ref{tab:LL_MA_test} shows the LL on the test data, covering the entire range of distances, for the four sub-models and for model averaging. As in estimation, we see that the model averaging structure obtains the highest LL. What is notable is the contrast with the results on the estimation data, i.e. Table \ref{tab:LL_estimation}. For both datasets, the gains achieved by model averaging relative to the best-performing sub-model are larger on the test data than on the estimation data. 

\begin{table}[h]
    \centering
    \vspace{0.5cm}
    \caption{Log-Likelihood for model averaging: individual components and overall structure on test data (covering distance segments 1-10), with gains by MA over individual models shown in brackets}
    \label{tab:LL_MA_test}
    \begin{tabular}{rcc}
        \toprule
        Model & \multicolumn{1}{c}{DECISIONS} & \multicolumn{1}{c}{LPMC} \\
        \midrule
    NL/MNL    & -825.88 (7.14\%) & -11,764.56 (5.52\%) \\
    DFT   & -805.78 (4.82\%) & -11,426.72 (2.72\%) \\
    MLP   & -804.91 (4.72\%) & -11,448.70 (2.91\%) \\
    XGB   & -806.62 (4.92\%) & -12,037.51 (7.66\%) \\
    \textbf{MA}    & \textbf{-766.88} & \textbf{-11,115.43} \\
    \bottomrule
    \end{tabular}
    \vspace{0.5cm}
 \end{table}

There are two possible influences that could drive the differences between Table \ref{tab:LL_MA_estimation} and Table \ref{tab:LL_MA_test}. The first of these is the potential that model averaging reduces the risk of over-fitting to the estimation data, which can be tested by comparing the performance on segments 2-9 between the estimation and test data. The second is the potential that model averaging offers improved out-of-distribution performance, which can be tested by comparing the performance on segments 1 and 10 between the sub-models and the model averaging structure.

\begin{sidewaysfigure}[hp!]
\includegraphics[clip, trim=2cm 0 2cm 0, height=5cm]{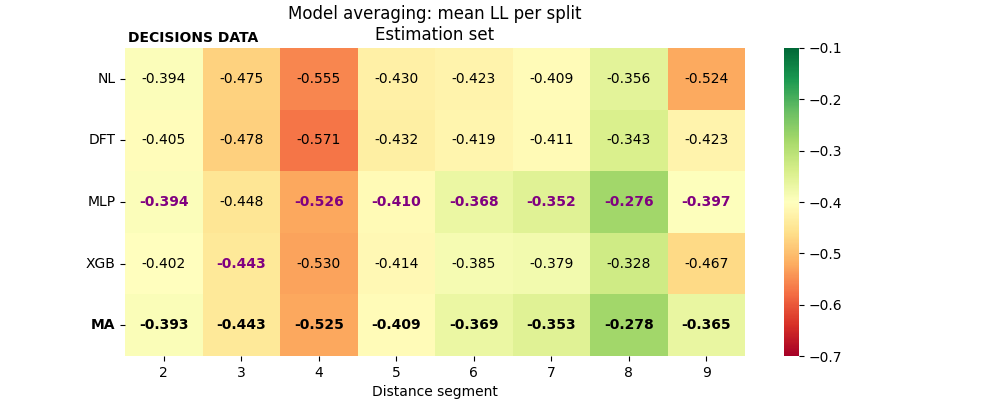}
\includegraphics[clip, trim=2cm 0 2cm 0, height=5cm]{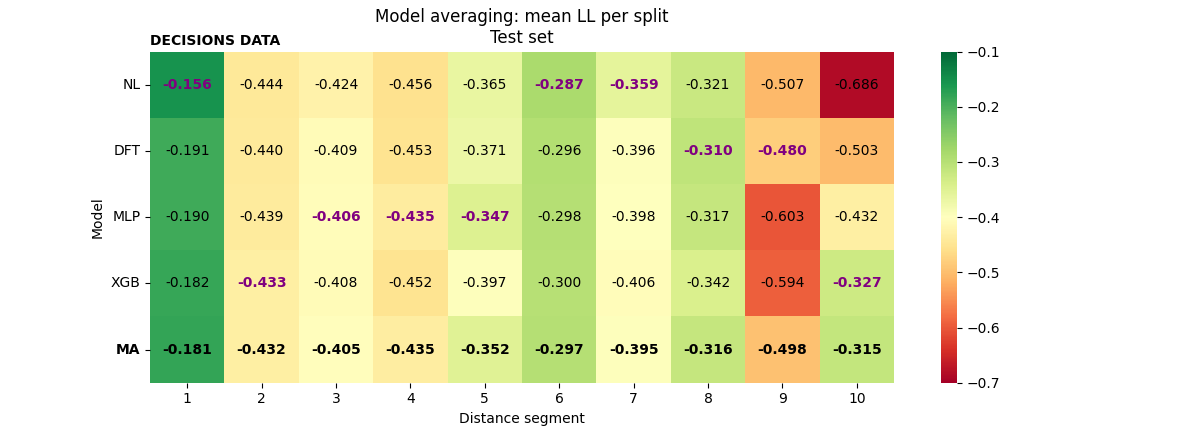}\\
\includegraphics[clip, trim=2cm 0 2cm 0, height=5cm]{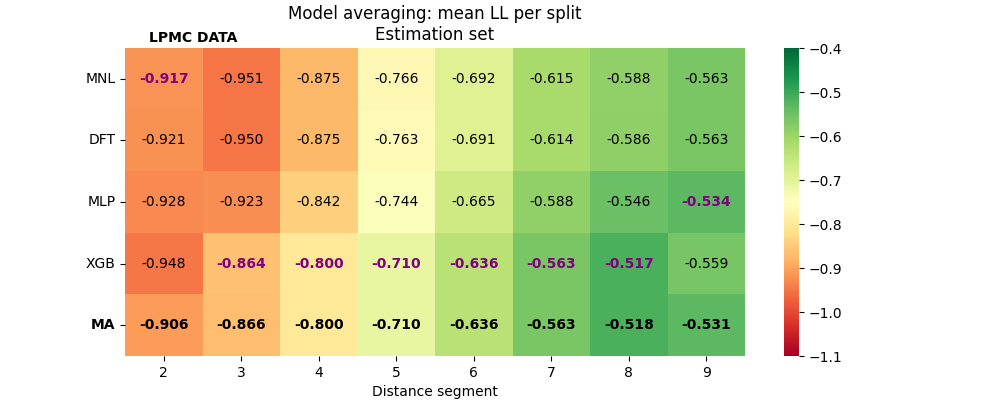}
\includegraphics[clip, trim=2cm 0 2cm 0, height=5cm]{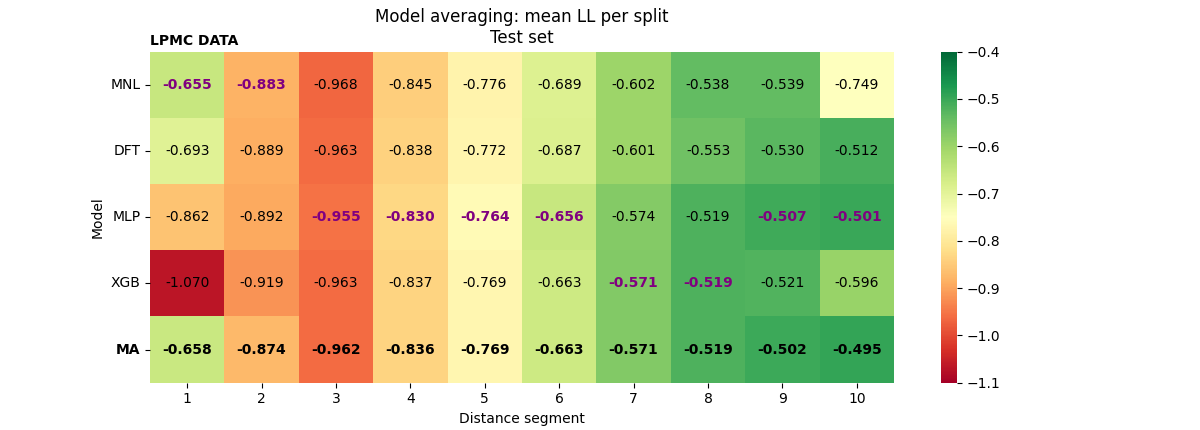}
\caption{MA Estimation and test performance}	
\label{fig:est_val_heat}
\end{sidewaysfigure}

To support this discussion, Figure \ref{fig:est_val_heat} shows the per observation log-likelihoods in the different segments for estimation and test. We first see that, in estimation, model averaging outperforms (or equals) the four sub-models in 5 out of 8 estimation segments for the DECISIONS data, and 6 out of 8 for the LPMC data. When we turn to the test data, model averaging performs best (or joint best) in 4 out of 10 segments for the DECISIONS data, and 5 out of 10 segments for the LPMC data. These per-segment statistics go hand in hand with the best overall LL obtained by model averaging as reported in Table \ref{tab:LL_MA_estimation} and Table \ref{tab:LL_MA_test}.

We earlier raised two possible reasons why the advantages of MA might be larger on the test data than on the validation data. In relation to over-fitting to the estimation data, we see that, for all four sub-models as well as for the model averaging structure, there are segments (out of 2-9) where the per observation LL increases and segments where it decreases. No clear pattern emerges in the comparison across models and segments. Turning to the second potential reason, namely out-of-distribution performance, we observe that for the DECISIONS data, model averaging is outperformed by NL in segment 1, but obtains the best performance in segment 10, where its combined performance on segments 1 and 10 is higher than for any of the other models ($-0.236$ \emph{vs} $-0.241$ for XGB). For the LPMC data, a similar story emerges. Model averaging is outperformed slightly by NL in segment 1, but obtains the best performance in segment 10. Again, overall, MA performs the best of all models when looking at segments 1 and 10 together ($-0.574$ \emph{vs} $-0.601$ for DFT). 

Overall, these findings lead to the conclusion that there is little difference across the models in terms of out-of-sample performance when looking at segments 2-9. On the other hand, there is clear evidence that our model averaging approach offers benefits in terms of out-of-distribution performance, which was its intended aim.

\section{Conclusions}\label{sec:conclusions}

This work has taken an important step forward in combining insights from different modelling approaches for travel demand forecasting. Specifically, we have shown that different models predict choices differently well whether we are within the ranges of the estimation data, or outside. This result is not surprising in itself but is quantified by our work.

The main contribution comes in the development of an MLP-based model averaging approach that estimates weights for different models as a function of the \emph{distance} away from the estimation data. Using trip distance as the metric to characterise a trip, across two different case studies looking at mode choice, we see that data-driven machine learning approaches obtain almost the entire weight when making predictions for trips in \emph{areas} similar to the trips in the estimation data (i.e. similar distances). However, once we move outside that area and look at out-of-distribution prediction, the model averaging approach increases the weight for the models with behavioural underpinnings, sometimes substantially so. 

In terms of prediction performance, we observe that model averaging offers the best overall performance on both the estimation and test data. Crucially, this advantage is at its largest when making out-of-distribution predictions, in our case, predicting mode choice for trips whose distance is outside the range covered by the estimation data.

While the analysis focuses on predictive performance, the observed differences across model classes can be interpreted in light of their underlying structure. In particular, models with a stronger behavioural underpinning may exhibit more stable performance in out-of-distribution settings because their parameters are constrained by behavioural theory and regularity conditions. This can limit the extent to which such models capture spurious patterns or context-specific correlations present in the estimation data. In contrast, more flexible, data-driven models may achieve superior in-sample fit by capturing patterns that do not generalise, but this flexibility can lead to reduced robustness when extrapolating beyond the observed data. We emphasise that this interpretation remains tentative and is not directly tested in this study, but it provides a plausible explanation for the observed performance differences and aligns with broader discussions on the trade-off between flexibility and generalisation.

Our work relies on multiple separate models, and this clearly leads to increased complexity as well as doing nothing to address the black box nature of machine learning approaches. A contrasting approach is to make the trade-off between fit, generalisation, and interpretability more explicit at the estimation stage. Recent work has started to do so by proposing multi-objective estimation frameworks that jointly consider in-sample fit and out-of-sample predictive accuracy, thereby making it possible to explore Pareto-efficient specifications rather than selecting models on fit alone \citep{beeramoole2024multiobjective}. Related work has also shown that machine learning methods can be made more behaviourally credible by embedding behavioural regularity or interpretability constraints, which may help improve robustness when moving beyond the estimation domain \citep{feng2024gradient}. At the same time, the recent benchmarking literature makes clear that choice modellers are right to take the predictive strengths of machine learning seriously: across a very large empirical comparison, many ML models outperform traditional discrete choice models in prediction accuracy, even while important questions remain around transferability and uncertainty \citep{wang2024benchmark}. It would therefore be unwise to disregard the strong in-sample performance that data-driven methods can deliver. Instead, our model averaging approach is motivated precisely by the idea of combining the strengths of both traditions: the behavioural interpretability and theoretical discipline of choice models, and the empirical flexibility and predictive power of machine learning methods.

There are, of course, some opportunities for further refinement. The key one relates to the fact that, in applying our proposed model averaging approach, an analyst needs to define what attribute(s) to use to characterise to what extent a trip is \emph{within distribution} or \emph{out-of-distribution} means. In our work, we have focused on the obvious example of trip distance, for which there were pragmatic reasons, but also given that trip distance embodies key information in relation to similarity between trips. A core opportunity for future work is thus the investigation of out-of-distribution performance considering attributes besides trip distance alone, for example, going outside the range of socio-economic characteristics covered by the estimation data, or combinations of socio-economic characteristics not covered in the estimation data. Out-of-distribution could also take on a multivariate nature, thus covering multiple trip characteristics at the same time. This, however, then also necessitates care by the analyst in defining what constitutes within-distribution and out-of-distribution. We take initial steps in that direction in Appendix \ref{sec: Appendix B}.

Finally, we employed an MLP as the meta-model within the model averaging framework. The literature on ensemble and stacking architectures in machine learning offers a broad range of methodological extensions that could further enhance such meta-modelling approaches \citep[e.g.][]{shaikh2024fundamental}. Future research could, for instance, investigate deep stacking networks, Bayesian ensemble formulations, or ensemble distillation methods to improve model robustness and generalisation, particularly in data-scarce settings. Furthermore, in cases where a larger number of explanatory variables are available and one wishes to understand their importance, more structured meta-models, such as a logistic regression model or hybrid models, such as TasteNet \citep{HAN2022166} can be explored.  

\section*{Acknowledgements}

Stephane Hess acknowledges the support of the Delft Excellence Fund, as well as the European Research Council through the advanced grant 101020940-SYNERGY. Sander van Cranenburgh acknowledges the support of the TU Delft AI Labs programme. The authors are thankful to Thomas Hancock, Tim Hillel and Georges Sfeir for comments on an earlier draft. The comments of three anonymous referees have helped us to substantially improve the paper.

\appendix

\section{Cross-validated stacking LPMC}\label{sec: Appendix A}
To assess whether the absence of a separate validation set for model averaging affects the results, we conducted an additional analysis using the larger LPMC dataset in which a dedicated split is introduced.

First, the base models (level-0 models) are trained exclusively on the estimation data. These trained models are then applied to a separate validation set that was not used during the training of the base models. The predictions generated by the base models on this validation set form the inputs for the model averaging stage.

Next, the model averaging MLP (level-1 model) is trained using this validation data. To facilitate training and prevent overfitting, the validation set is further split into a \texttt{val\_train} and \texttt{val\_test} subset. The \texttt{val\_train} subset is used to estimate the parameters of the model averaging MLP, while the \texttt{val\_test} subset is used to monitor predictive performance during training. Training is stopped early when performance on the \texttt{val\_test} subset deteriorates, implementing an early stopping criterion.

This procedure ensures that the level-1 model is estimated using predictions obtained from data that were not used to train the base models, thereby providing a stricter separation between training stages. While this procedure is not identical to cross-validated stacking as typically implemented in the stacking literature, it closely mirrors the workflow used in the main analysis. As such, it provides a meaningful assessment of whether the absence of a dedicated validation set for model averaging affects the results reported in the main body of the paper.

\begin{figure}[h]
\centering
\includegraphics[width=15cm]{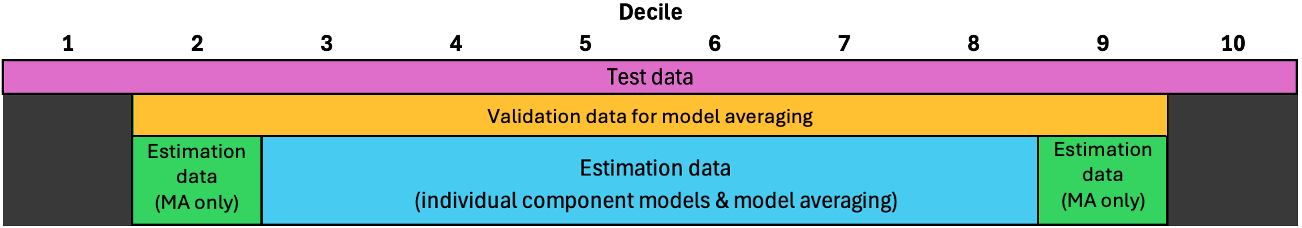}\\
\caption{Data splitting with validation}	
\label{fig:data_split_with_val}
\end{figure}

The resulting splitting is illustrated in Figure \ref{fig:data_split_with_val}. 

\begin{figure}[htbp]
\centering
\includegraphics[width=\textwidth]{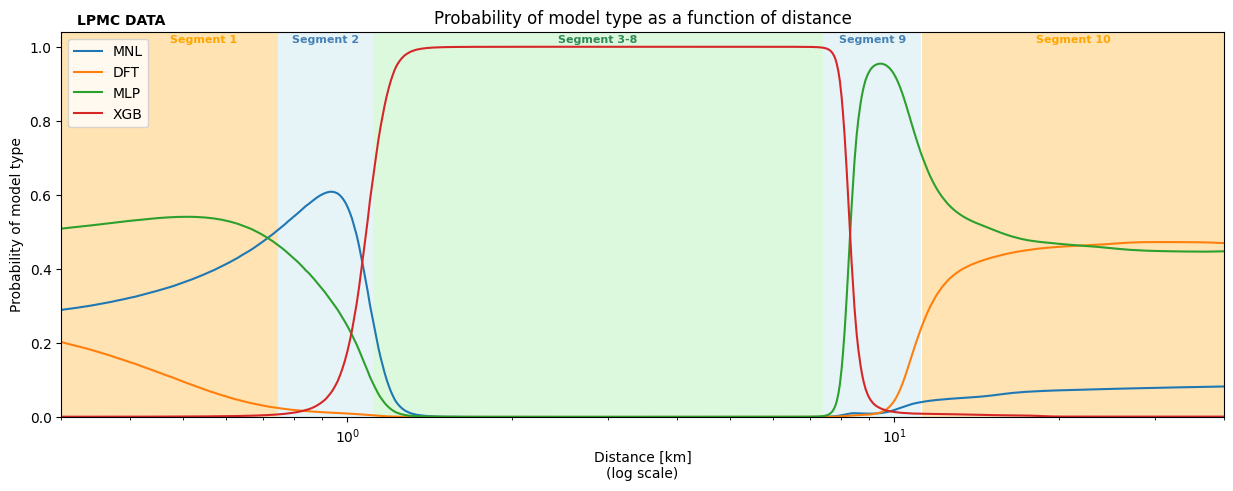}
\caption{Model averaging weights as a function of distance for the LPMC data using the validation-based stacking procedure}
\label{fig:MA_LPMC_appendix}
\end{figure}

Below, we present the main results to facilitate comparison with those reported in the paper. The fit of the model averaging approach on the test data (covering distance segments 1-10) remains largely unchanged, with a log-likelihood of −11,204.87, vs. -11,115.43 in the paper, implying a drop in LL by $0.8\%$. More importantly, Figure \ref{fig:MA_LPMC_appendix} shows broadly the same mixed pattern as a function of distance. One difference is that the allocation of weights across models is somewhat sharper and less gradual. In particular, within distance segments 3–8 virtually all weight is assigned to XGB, whereas outside this range XGB receives almost no probability mass. In segment 9, MLP attains the largest share of the probability, while segments 1 and 10 exhibit a mixture of MNL, DFT, and MLP.

\begin{figure}[htbp]
\centering
\includegraphics[width=\textwidth]{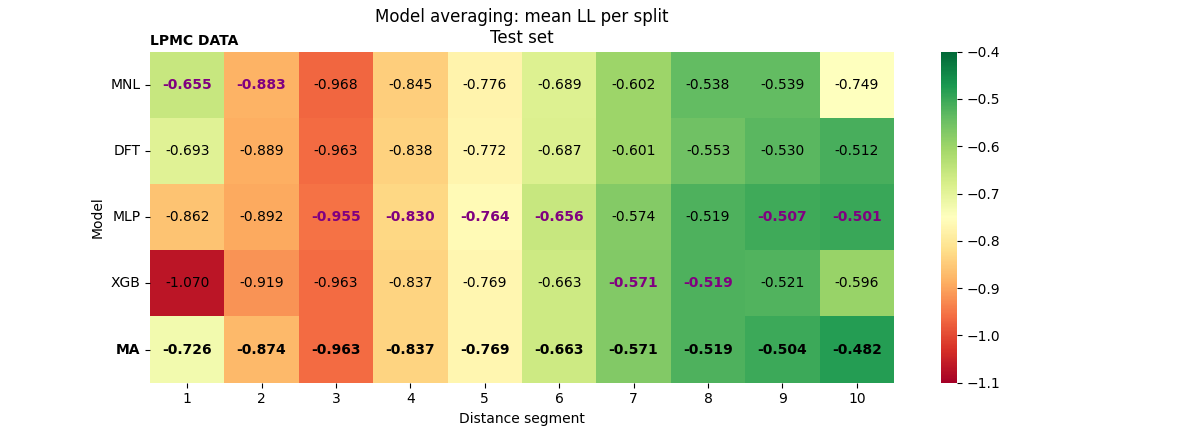}
\caption{Heatmap of model performance across distance segments}
\label{fig:heatmap_test_LPMC_appendix}
\end{figure}

A comparison of Figure \ref{fig:heatmap_test_LPMC_appendix} with the results reported in the paper reveals a largely similar pattern. The numerical values change only marginally. The only noteworthy difference is a modest increase in the mean log-likelihood achieved by model averaging in distance segment 1, which rises to $-0.726$ (vs. $-0.658$ in the paper). The slightly sharper allocation may be related to the revised estimation procedure, where the level-0 models and the meta-model are trained on separate data partitions following \cite{hastie2009elements}. Training the meta-model on predictions that reflect out-of-sample performance can affect the allocation of weights across models. However, given the very modest changes observed, the results suggest that not using a separate validation set for model averaging has little impact on our overall findings.

\section{Illustration using multivariate measure for out-of-distribution scale}\label{sec: Appendix B}

The main body of the paper focuses on a single variable to compute a measure of how in-distribution or out-of-distribution a given observation is. In this appendix, we provide an illustration of how multivariate measures can be used, with a focus on the DECISIONS data.

When using multiple variables to create a measure of out-of-distribution, an analyst first needs to decide whether those metrics are used as separate variables, thus providing multiple inputs into the weight function for model averaging (cf. Eq. \ref{eq:MA_likelihood_obs}), or whether they are first combined into a single metric. In the implementation in this appendix, we use the latter approach, and the contrast between the two remains an avenue for future work. The application in the main body of the paper already showed the ease of using multiple variables in the model averaging weigthing function (in the form of different transformations of the distance variable).

A second decision relates to how to segment the data, i.e. the creation of 
$D_{train,sub}$ and $D_{train,MA}$. If the metric is transformed into a single variable, then the same approach as was used Section \ref{sec:data_splitting} can be used. If multiple variables are retained at the same time for including in Eq. \ref{eq:MA_likelihood_obs}, then the data splitting can still be conducted on the basis of a combined metric, making observations more likely to be seen as out-of-distribution if they fall out-of-distribution for multiple variables. In the example in the present appendix, we rely on the same splits as in \ref{sec:data_splitting} for ease of comparison. Again, further investigation of this remains an avenue for future work.

\subsection{Multivariate metric}

Multiple approaches exist for constructing out-of-distribution (OOD) metrics, ranging from density-estimation methods and nearest-neighbour measures to distance-based metrics. In the present application, we require a measure that can accommodate the diverse set of variables available in travel behaviour data, including continuous, ordinal, and categorical variables. We therefore use the Gower distance \citep{gower1971general}, which is among the most widely used distance measures for mixed-type data. Its principal advantage is that it provides a common scale for comparing observations across variables of different types, allowing a single multivariate measure of similarity to be constructed. Following standard extensions of the original formulation, ordinal variables are transformed to ranks prior to calculating pairwise dissimilarities. We then use the resulting pairwise Gower distances to quantify how unusual an observation is relative to a reference population.

Let $x_{nk}$ denote the value of variable $k$ for observation $n$. For a pair of observations $n$ and $m$, the Gower distance is defined as:

\begin{equation}\label{eq:gower}
d_{nm} = \frac{\sum_{k=1}^{K} w_{nmk}\,\delta_{nmk}}
{\sum_{k=1}^{K} w_{nmk}},
\end{equation}

\noindent where $\delta_{nmk}$ is the variable-specific dissimilarity measure and $w_{nmk}$ is an indicator taking the value one when the comparison is valid and zero otherwise:

\begin{equation}
w_{nmk}
=
\begin{cases}
1 & \text{if variable } k \text{ is applicable for both } n \text{ and } m,\\
0 & \text{otherwise}.
\end{cases}
\end{equation}

\noindent As a result, observations are compared only on dimensions that are relevant to both observations, preventing non-applicable characteristics from artificially increasing the measured dissimilarity.
In some cases, a value of zero indicates that a variable is not applicable rather than representing a genuine observed value. Such structural zeros are recoded as missing values prior to the calculation of the Gower distance. Through the weighting term $w_{nmk}$, these variables are excluded from the pairwise comparison. Neither of these cases arises with our data, but this is noted here for completeness.

For continuous variables, the variable-specific dissimilarity is computed as:

\begin{equation}
\delta_{nmk}
=
\frac{|x_{nk}-x_{mk}|}
{\max(x_k)-\min(x_k)},
\end{equation}

\noindent where the denominator corresponds to the observed range of variable $k$ in the reference sample. For nominal categorical variables,

\begin{equation}
\delta_{nmk}
=
\begin{cases}
0 & \text{if } x_{nk}=x_{mk},\\
1 & \text{otherwise}.
\end{cases}
\end{equation}

\noindent Ordinal variables are first transformed to ranks and rescaled to the unit interval. Let $r_{nk}$ denote the rank of observation $n$ on variable $k$, and let $G_k$ denote the number of ordered categories. The transformed value is:

\begin{equation}
z_{nk}
=
\frac{r_{nk}-1}
{G_k-1},
\end{equation}

\noindent and the corresponding dissimilarity is:

\begin{equation}
\delta_{nmk}
=
|z_{nk}-z_{mk}|.
\end{equation}

\noindent The overall Gower distance is then obtained by averaging the variable-specific dissimilarities across all variables for which a valid comparison is available. The resulting distance measure lies on a common scale and can therefore be applied to datasets containing a mixture of variable types.

This approach thus combines multiple variables into a single metric. If an analyst wishes to compute a separate metric for each variable, and have multiple variables enter into Eq. \ref{eq:MA_likelihood_obs}, then Eq. \ref{eq:gower} can simply be calculated separately for each variable, thus not summing over $K$. This would allow the weights of different variables to be estimated separately, i.e. showing different impacts of being out-of-distribution for each variable.

From the distance metric, we now need to calculate an out-of-distribution measure. Let $\mathcal{R}$ denote the set of observations used to define the reference distribution, in our case the data used in estimation of the individual models. For each observation $n$, we calculate an out-of-distribution (OOD) score as the average distance to all observations in the reference sample:

\begin{equation}
s_n = \frac{1}{|\mathcal{R}_n|}
\sum_{m \in \mathcal{R}_n} d_{nm},
\end{equation}

\noindent where $\mathcal{R}_n = \mathcal{R}\setminus\{n\}$ when observation $n$ itself belongs to the reference sample, and $\mathcal{R}_n = \mathcal{R}$ otherwise. Larger values of $s_n$ indicate that an observation is, on average, further away from the reference population and therefore less representative of the patterns observed in the estimation data.

To facilitate interpretation, the raw OOD scores are standardised using the distribution of scores observed within the reference sample. Let $\bar{s}_{\mathcal{R}}$ and $\sigma_{\mathcal{R}}$ denote the mean and standard deviation of $s_n$ across all observations in the reference sample, $\mathcal{R}$. The standardised OOD score is then given by:

\begin{equation}
z_n =
\frac{s_n-\bar{s}_{\mathcal{R}}}
{\sigma_{\mathcal{R}}}.
\end{equation}

\noindent A value of $z_n=0$ corresponds to an observation whose average distance from the reference population is equal to the average observed within the reference sample. Positive values indicate observations that are more unusual than the typical reference observation, while negative values indicate observations that are more centrally located within the distribution. 

The above calculations can be carried out for observations that are in the original sample as well as new observations. Indeed, once $\bar{s}_{\mathcal{R}}$ and $\sigma_{\mathcal{R}}$ have been estimated, the same transformation can be applied to observations that were not used in the construction of the reference distribution. For such observations, the score $s_n$ is obtained by averaging the Gower distance to all observations in the reference sample and then standardising using the reference-sample mean and standard deviation. If these new observations differ systematically from the estimation sample, their average distances will tend to be larger, resulting in more positive values of $z_n$. The resulting measure therefore provides a direct indication of how far an observation lies from the empirical distribution represented by the estimation sample, while remaining directly comparable across estimation and holdout samples.

\subsection{Implementation for DECISIONS data}

We examined a multivariate measure of similarity based on Gower distance for the DECISIONS dataset. This is an illustration to show how the model averaging approach can be based on a multidimensional out-of-distribution measure, but without an attempt to fine tune the approach. To ensure direct comparability with the results reported in the main body of the paper, we retain exactly the same training-test partition as in the primary analysis, i.e. including the splitting into distance segments. Consequently, the multivariate measure is not used to construct the partition itself; it is only used as an alternative OOD input to the model-averaging weight function. This is likely to provide a conservative illustration of the potential benefits of a multivariate approach, since the partition was not constructed to maximise separation in the multivariate space.

In our example, the multivariate representation incorporates level-of-service information (e.g. time and cost), mode availability, socio-demographic (e.g. age and income) and geographic information (e.g. distance). This thus captures differences across a broad range of dimensions simultaneously, allowing us to objectively quantify out-of-distribution. The multivariate representation is constructed in part using the same explanatory variables that define the behavioural context in which the constituent models are applied. The objective is not to re-estimate the models, but rather to characterise the location of each observation within the explanatory-variable space and thereby assess the similarity between observations. This information can subsequently be used to determine the relative weights assigned to the constituent models when generating averaged predictions. As the constituent models and the model averaging structure are not estimated simultaneously, this double use does not in itself create a circularity issue

For each observation, we compute an out-of-distribution (OOD) score based on the average Gower distance to all observations in the training sample. These scores are then standardised using the mean and standard deviation observed within the training sample. By construction, the average OOD score in the training sample is approximately zero. Applying the same standardisation to the test sample yields an average OOD score of 0.884, indicating that observations in the test sample are, on average, almost one standard deviation further away from the training-sample distribution than the training observations themselves.

Table~\ref{tab:gower_segments} reports the average OOD score by distance segment and sample. Several observations emerge. First, the most extreme distance segments exhibit the largest positive OOD scores. In particular, segments 1 and 10 have average scores of 1.25 and 2.20 respectively, while segments 2 and 9 also show substantially positive values. While the multivariate OOD score was not used in constructing the partition, the most extreme distance segments exhibit substantially larger OOD scores than the central segments. This indicates that the original distance-based partitioning also induces meaningful separation in the multivariate explanatory-variable space. Second, the central distance segments generally exhibit negative or near-zero scores, indicating that these observations are more representative of the multivariate distribution observed in the training sample. Third, within the distance segments that contain both training and test observations, OOD scores are broadly comparable, suggesting that the multivariate measure is primarily capturing differences between distance segments rather than differences arising from the allocation of observations to training and test samples within a segment.

\begin{table}[ht]
\centering
\caption{Average multivariate OOD score by distance segment}
\label{tab:gower_segments}
\begin{tabular}{lcc}
\toprule
Distance segment & Training & Test \\
\midrule
1  & --      & 1.252 \\
2  & --      & 0.665 \\
3  & 0.316   & 0.032 \\
4  & 0.010   & -0.317 \\
5  & -0.141  & -0.509 \\
6  & -0.187  & -0.441 \\
7  & -0.133  & -0.243 \\
8  & 0.131   & -0.088 \\
9  & --      & 0.808 \\
10 & --      & 2.195 \\
\midrule
Overall & 0.000 & 0.884 \\
\bottomrule
\end{tabular}
\end{table}

These findings are noteworthy because the original partitioning was based exclusively on a univariate distance measure, whereas the Gower-based score is derived from a joint distribution spanning approximately thirty explanatory variables. The fact that the observations identified as extreme under the distance-based segmentation also tend to exhibit substantially larger multivariate OOD scores suggests that the original partitioning induces meaningful separation in the broader explanatory-variable space. While this exercise should not be interpreted as a fully optimised application of the multivariate approach, it nevertheless demonstrates that a Gower-based measure is capable of identifying observations that differ systematically from the estimation sample and could therefore be used directly to construct transferability partitions in future applications. In fact, we could expect larger benefits in such cases than we observe here, where the data splitting relied on a single variable which is only one of the constituent members of the Gower measure. 

\subsection{Results for DECISIONS data}

We start by looking at the results obtained on the estimation data for model averaging, covering segments 2-9, but with model parameters for the individual models estimated on segments 3-8 only. 

Table \ref{tab:LL_MA_estimation_OOD} is the analogue of \ref{tab:LL_MA_estimation} from the main body of the paper. The fit of the MA model is slightly inferior to what we saw in the main body of the paper, at $-2,769.66$ rather than $-2,752.68$. This however is not surprising as the data splitting was geared towards the use of distance alone. Even so, the MA structure still clearly obtains a higher LL than any of the four individual models. 

\begin{table}[h]
    \centering
    \vspace{0.5cm}
    \caption{Log-Likelihood for model averaging on DECISIONS data using Gower measure for MA (covering distance segments 2-9), with gains by MA over individual models shown in brackets}
    \label{tab:LL_MA_estimation_OOD}
    \begin{tabular}{rc}
        \toprule
        Model & \multicolumn{1}{c}{DECISIONS} \\
        \midrule
    NL & \multicolumn{1}{l}{-3,145.33 (11.94\%)}\\
    DFT    & \multicolumn{1}{l}{-3,063.46 (9.59\%)} \\
    MLP    & \multicolumn{1}{l}{-2,787.34 (0.63\%)}\\
    XGB    & \multicolumn{1}{l}{-2,949.47 (6.10\%)} \\
    \textbf{MA} & \textbf{-2,769.66} \\
        \bottomrule
    \end{tabular}
    \vspace{0.5cm}
\end{table}

\noindent We next look at the performance of the model averaging structure on the test data. Table \ref{tab:LL_MA_test_OOD} is the counterpart of \ref{tab:LL_MA_test}, showing the LL on the test data, covering the entire range of distances, for the four sub-models and for model averaging. The fit of the MA model is again slightly less than in the main body of the paper, at $-771.42$ vs $-766.88$, but MA still remains the best fitting model, and the gains by model averaging over the best-performing sub-model are again larger on the test data than on the estimation data. 

\begin{table}[h]
    \centering
    \vspace{0.5cm}
    \caption{Log-Likelihood for model averaging on DECISIONS data using Gower measure for MA (covering distance segments 1-10), with gains by MA over individual models shown in brackets}
    \label{tab:LL_MA_test_OOD}
    \begin{tabular}{rcc}
        \toprule
        Model & \multicolumn{1}{c}{DECISIONS} \\
        \midrule
    NL/MNL    & -825.88 (6.60\%) \\
    DFT   & -805.78 (4.27\%) \\
    MLP   & -804.91 (4.16\%) \\
    XGB   & -806.62 (4.36\%) \\
    \textbf{MA}    & \textbf{-771.42} \\
    \bottomrule
    \end{tabular}
    \vspace{0.5cm}
 \end{table}

Finally, in the main body of the paper, we reported that MA obtained the best average LL per observation in segments $1$ and $10$, at $-0.236$ \emph{vs} $-0.241$ for XGB. With the new MA model, the performance in fact improves further in those segments, now reaching an average of $-0.222$. This thus provides further support as to the benefit of the approach, including when using a multivariate metric. The fact that these advantages arise even though the data was split by distance alone is especially encouraging.

\bibliographystyle{elsarticle-harv}
\bibliography{references}

\end{document}